# Rotational control of asymmetric molecules: dipole- vs. polarizability- driven rotational dynamics

Ran Damari[1], Shimshon Kallush[2,3†] and Sharly Fleischer[1‡]

[1]*Raymond and Beverly Sackler Faculty of Exact Sciences, School of Chemistry, Tel Aviv University, Tel Aviv 6997801, Israel*
[2]*Dept. of Physics and Optical Engineering, ORT Braude College, PO Box 78, 21982 Karmiel, Israel.*
[3]*The Fritz Haber Research Center and The Institute of Chemistry, The Hebrew University, Jerusalem 91904, Israel*

Abstract:

We experimentally study the optical- and terahertz- induced rotational dynamics of asymmetric molecules in the gas phase. Terahertz and optical fields are identified as two distinct control handles over asymmetric molecules, as they couple to the rotational degrees of freedom via the molecular- dipole and polarizability selectively. The distinction between those two rotational handles is highlighted by different types of quantum revivals observed in long duration (>100ps) field-free rotational evolution. The experimental results are in excellent agreement with Random Phase Wave Function simulations [Phys. Rev. A 91, 063420 (2015)] and provide verification of the RPWF as an efficient method for calculating asymmetric molecular dynamics at ambient temperatures, where exact calculation methods are practically not feasible. Our observations and analysis pave the way for orchestrated excitations by both optical and THz fields as complementary rotational handles, that enable a plethora of new possibilities in three-dimensional rotational control of asymmetric molecules.

---------------------------------------------------------------------------------------------------

## Introduction

Non-ordered media such as gas, liquid and most solid phases consist of a large number of individual molecules oriented in all possible directions. Such samples are termed 'isotropic' and lack any angular association between the molecular and the laboratory frames. As the interaction between light and molecule depends on the projection of the light polarization onto the dipole moment of the molecule, spectroscopic signals obtained from isotropic samples are inherently angularly averaged, impeding the extraction of various molecular-frame dependent features of interest. Motivated by the need for lifting this angular averaging, vast efforts have been put in the last decades into controlling the angular distributions of gas phase molecular ensembles to enable ultrafast molecular dynamics and molecular-frame spectroscopies in general.

Most of the efforts toward anisotropic angular distributions have focused on molecular *alignment* in which the most polarizable molecular axis align with a chosen lab-frame axis (typically referred to as the lab-frame z-axis). Recently, angular control efforts have gradually shifted toward molecular *orientation* in which the permanent dipoles of the molecules point preferentially in a chosen lab-frame direction (positive or negative lab-frame z-axis). Oriented samples provide yet an important feature that is absent from aligned molecular samples - the transiently lifted inversion symmetry of oriented samples provide the necessary condition for various nonlinear phenomena governed by even-orders in the light's electric field such as even harmonic generation[1], directional molecular ionization/dissociation[2] and $\chi^{(2n)}$-dependent optical signals in general. Throughout the years, angular coherent control explorations have yielded tremendous progress not only in the realm of molecular frame spectroscopy, but also in various tangential fields and applications (for recent review articles see [3,4,5]).

Our aims in this work are three-fold: to study and compare between optical- and THz- induced rotational dynamics in asymmetric molecules. To establish single-cycle THz fields as a dipole-selective rotational handle, complementary to ultrafast optical pulses. To provide experimental verification of the RPWF method (recently developed by us for this task) as a calculation method for rotational dynamics of asymmetric molecules at ambient temperatures.

Before delving into the dynamics of asymmetric tops we note that the vast majority of ultrafast rotational control have focused on linear molecules and symmetric tops that are typically modeled as quantum mechanical rigid rotors with effectively only one moment of inertia. Molecules belonging to this group show perfect (up to centrifugal distortion) recurrences manifested by periodic anisotropic distributions with each revival time, given by $T_{rev}=1/2Bc$ with $B$ the rotational coefficient in [cm$^{-1}$] and $c$ the speed of light [6,7]. Moreover, most of the control efforts have focused on molecular alignment that is induced via the polarizability components of the molecules. Significantly fewer works have focused on the orientation of molecules, and even fewer have focused on the orientation of asymmetric molecules, which is the at the focus of this work, motivated by the fact that practically all of the molecules of interest to chemistry and biology belong to the group of asymmetric molecules. In addition to their complicated quantum rotational dynamics, asymmetric rotors introduce highly demanding computational challenges [8, 9, 10, 11] as will be described later in the text.

**Laser induced rotations of asymmetric tops**

The theoretical and experimental foundations of asymmetric-top rotational dynamics have been laid by Felker [12] more than two decades ago, and have been actively reinforced since [13,14,15,16,17,18,19,20,21]. Alignment of asymmetric tops have been achieved by elliptically polarized pulses [18] and by multiple- time delayed pulses with well controlled polarizations [22]. Existing techniques for the orientation of asymmetric molecules rely on a combination of weak electrostatic fields (dc) and intense non-resonant pulses (near-IR) [23, 24 2525]. However, these techniques are valid for low rotational temperatures achieved via supersonic beam expansion and correspondingly low gas densities that prohibit all-optical spectroscopy such as optical birefringence measurements, high-harmonic generation spectroscopy etc. Moreover, the existence of permanent dc field is typically not favorable since orientation is not attained under field-free conditions. Other approaches for field-free molecular orientation have been demonstrated in linear molecules by either mixed field ($\omega+2\omega$) excitations [26,27] or single-cycle terahertz fields [28, 29, 30, 31, 32]. Both techniques are applicable at ambient gas temperatures and high gas densities that allow all-optical probing.

In this work we utilize intense single cycle terahertz fields as dipole-selective rotational handles of asymmetric molecules to induce orientation and rotational responses that are considerably different from those induced by optical pulses. We present both the terahertz- and optical- induced dynamics of $SO_2$ serving as an asymmetric top molecular model and has been the subject of recent works in the field [10, 11, 33].

**Theoretical:**

Asymmetric molecules like $SO_2$ are described by three different moments of inertia corresponding to rotation around their three principal axes as shown in Fig.1. The rotational Hamiltonian for the asymmetric rotor is given by $\hat{H} = \hat{H}_0 + \hat{V}(t)$ where $\hat{H}_0 = \frac{\hat{L}_a^2}{2I_a} + \frac{\hat{L}_b^2}{2I_b} + \frac{\hat{L}_c^2}{2I_c}$ is the field-free rotational Hamiltonian and $\hat{V}$ the interaction term of the molecule with the external field. We study two kinds of rotational dynamics, induced by two different excitation fields: an ultrashort near-IR pulse (FWHM~100fs duration, $\omega_{optical} = 3.75 \cdot 10^{14} Hz$), and a single-cycle terahertz pulse (FWHM~1ps, $\omega_{THz} \sim 5 \cdot 10^{11} Hz$).

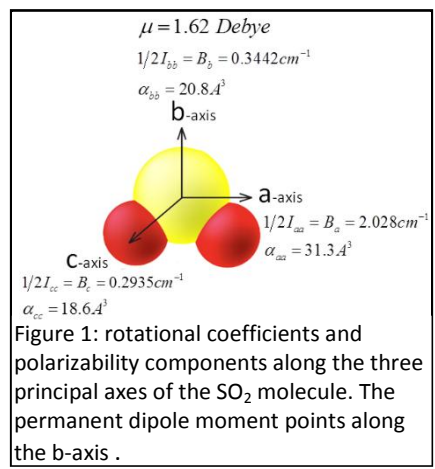

Figure 1: rotational coefficients and polarizability components along the three principal axes of the $SO_2$ molecule. The permanent dipole moment points along the b-axis.

**Excitation by an ultrashort near-IR (optical) pulse:**

At the optical frequency of 12,500 cm$^{-1}$ the laser field is far off-resonance from the rotational transition frequencies of molecules, in the few cm$^{-1}$ regime. In this case the dipole interaction term averages out after few oscillations of the fields, and the rotational dynamics is induced via the anisotropic polarizability components of the molecule (Fig.1). As was previously shown [11,12,15,10,9], the interaction with a linearly polarized optical field is given by $\hat{V}_{optical} = -\frac{1}{4}\varepsilon^2(t)[\alpha_{aa}\cos^2\theta_a + \alpha_{bb}\cos^2\theta_b + \alpha_{cc}\cos^2\theta_c]$ where E(t) is the pulse field envelope, $\alpha_{ii}$ is the polarizability component along the molecular frame ii-axis, and $\theta_i$ is the angle between the field polarization and the i$^{th}$ molecular frame-axis. The rotational dynamics is thus governed by the polarizability terms, with the main contribution from the most polarizable axis, the O-O axis of $SO_2$ (see Fig.1).

**Excitation by a single-cycle THz field:** With a frequency spanning the 3-30 cm$^{-1}$ range, the single-cycle THz field (~1ps duration) interacts resonantly with the molecules via their permanent dipoles and induce rotational motion [28,29

30,31,32]. The interaction term, $\hat{V}_{THz} = -\hat{\vec{\mu}} \cdot \vec{E}(t)$, rotates the dipole axis (b-axis of SO$_2$, see Fig.1) to orient in the direction of the THz field. Since the THz field oscillates much slower than the optical field, the molecular dipoles are able to follow the field throughout the excitation, namely the interaction is of resonant nature [29,34].

**RPWF simulation details**

Simulating the rotational dynamics of asymmetric top molecules is far more computationally demanding than linear or symmetric top molecules owing to the existence of three-coupled rotational degrees of freedom associated with the three different principal moments of inertia [8,9,10,11]. At ambient temperatures the large number of thermally occupied states makes exact calculation methods very costly in terms of computation time, and in many cases practically not feasible. For simulating the dynamics of asymmetric molecules at ambient temperatures we use the RPWF method, recently adopted by us for this purpose [35, 36, 9]. The initial state of the system is represented by rotational wavefunction, $\left| \vec{\theta}_k \right\rangle = \sum_{JM\tau} \sqrt{\frac{e^{-E_{JM\tau}/k_BT}}{Z}} e^{i\theta^k_{JM\tau}} \left| JM\tau \right\rangle$ with a Boltzmann weighted magnitudes for the wavefunctions but with randomly chosen phases $\theta^k_{JM\tau}$. Such an initial state contains coherences between the rotational states, that should be absent from the thermal ensemble. With increasing number of realizations (*k*), the randomly chosen phases and corresponding coherences cancel out due to destructive interferences and the correct, 'coherence-free' representation of the initial thermal ensemble is achieved [37,38]. We have recently shown that the convergence efficiency of the RPWF method to exact calculations increases with the initial temperature of the ensemble and with the exciting field strength, both increasing the size of the Hilbert space and thereby increasing the randomization of the superposition phases (for details see [9]).

The calculation of the optical-induced alignment of SO$_2$ is performed here by RPWF for the first time and provides yet another test bed for the integrity of the RPWF by comparison to the experimental results (Fig. 2).

For the simulated orientation of Fig. 4 we fed the RPWF simulation with the experimentally measured THz field (Fig. 4b) as the excitation field E(t) for the dipole interaction term $\hat{V}_{THz} = -\hat{\vec{\mu}} \cdot \vec{E}(t)$. In accordance to the spin symmetry selection rules, only rotational levels that are symmetric with respect to the *b*-axis were taken into account. Taking the molecular dipole axis as the z-axis of the molecule invoked the inclusion of rotational levels belonging to A and $B_z$ symmetry sub-groups, *i.e.*, only states with even K participate in the calculation [39].

Experimental:

The rotational responses of $SO_2$ to the optical and the THz fields were measured separately in two different experimental setups: a time-resolved optical birefringence setup and time-resolved THz-induced emission measurements briefly described hereafter. Both measurements were conducted on $SO_2$ gas at the pressure of 23-26torr.

**Optical induced birefringence in $SO_2$**

We start with the time-resolved birefringence measurements of the $SO_2$ gas sample at 25torr and ambient temperature, shown in Fig. 2. The experimental setup is similar to that recently reported in refs[40, 10] based on the weak-field polarization detection technique of Faucher and company [41,42].

A 100fs pulse is split to yield an optical pump (~50 µJ/pulse, 800nm) and a weak 400nm probe (second harmonic generated in a BBO crystal) linearly polarized 45° relative to the pump pulse polarization. The probe pulse is routed through a computer controlled optical delay stage, and recombined with the 800nm pump pulse by dichroic mirror, following which they collinearly propagate in the sample cell. The diameter of the pump beam is reduced by an iris to assure complete overlap with the probe throughout the sample cell and at the focus and to avoid laser-induced plasma in the gas. The polarization of the 400nm probe is analyzed by a $\lambda/4$ waveplate and a glan-taylor polarizer and detected by a single photodiode [43]. The experimental setup used for the optical induced birefringence measurements is presented in the Supplementary Information SI.1.

The observed transient birefringence signal is proportional to the change in the degree of alignment of the sample:

$$I_{Signal}(t) \propto [\alpha_{aa}(\langle \cos^2 \theta_a \rangle_{(t)} - 1/3) + \alpha_{bb}(\langle \cos^2 \theta_b \rangle_{(t)} - 1/3) + \alpha_{cc}(\langle \cos^2 \theta_c \rangle_{(t)} - 1/3)]$$

The signal consists of contributions from all three molecular axes' polarizabilities, $\alpha_{ii}$ (see Fig.1), weighted by the transient change in the alignment of the axes. The latter are given by the ensemble averaged projections of the principal molecular axes onto the polarization axis of the pump pulse (laboratory Z-axis), $\langle \cos^2 \theta_i \rangle_{(t)} - 1/3, \{i = a,b,c\}$.

The optical-induced birefringence of $SO_2$ in Fig.2 shows excellent agreement between the experimental (solid blue) and the RPWF simulated data (solid red). The dynamics consists of mainly the J-type and C-type revivals [12], with periods of $[2(B+C)]^{-1}$=26.1ps and $[4C]^{-1}$=28.4ps respectively. The weak birefringence oscillations observed starting from $t \sim 35 ps$ are also perfectly captured by the RPWF simulation (shown in the orange dashed frame in Fig. 2), thereby providing clear validation of the RPWF method for rotational dynamics.

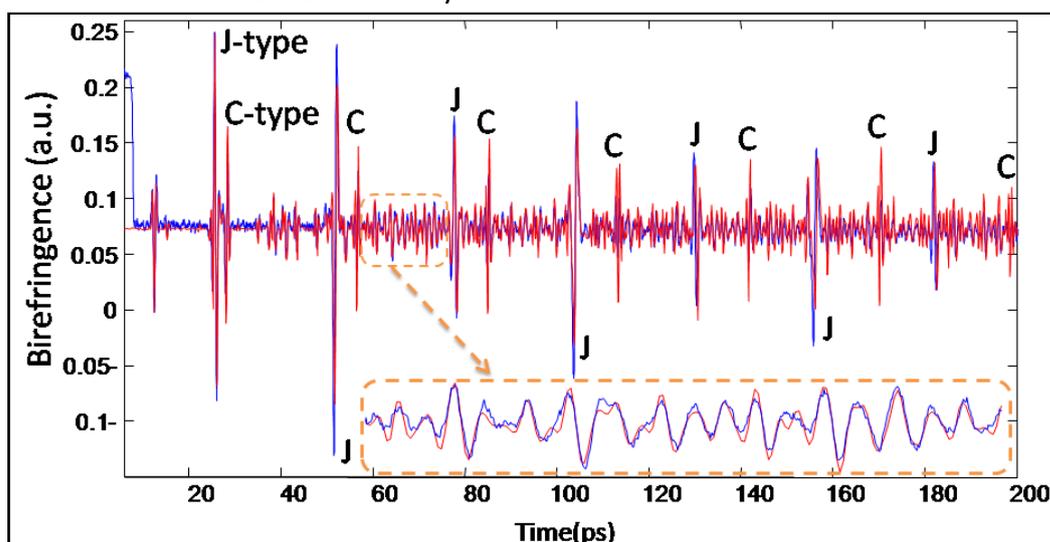

Figure 2: Experimental time-resolved birefringence (solid blue) and RPWF simulated alignment (solid red) from 25torr $SO_2$ gas sample at ambient temperature, following excitation by a 100fs optical pulse (800nm). The J-type revivals (with period of 26.1ps) and the C-type revivals (period 28.4ps) are marked in the figure. Note the alternating signal phases of the J-type with each revival and the constant phases of the C-type signals. The dashed orange inset depicts the weak oscillatory birefringence signals at the 58-76ps time frame, with close to perfect correspondence between the experimental and RPWF simulated data.

To examine the actual dynamics associated with these two types of revivals, we present in figure 3 the transient alignment of the three principal molecular axes onto the Z-lab axis.

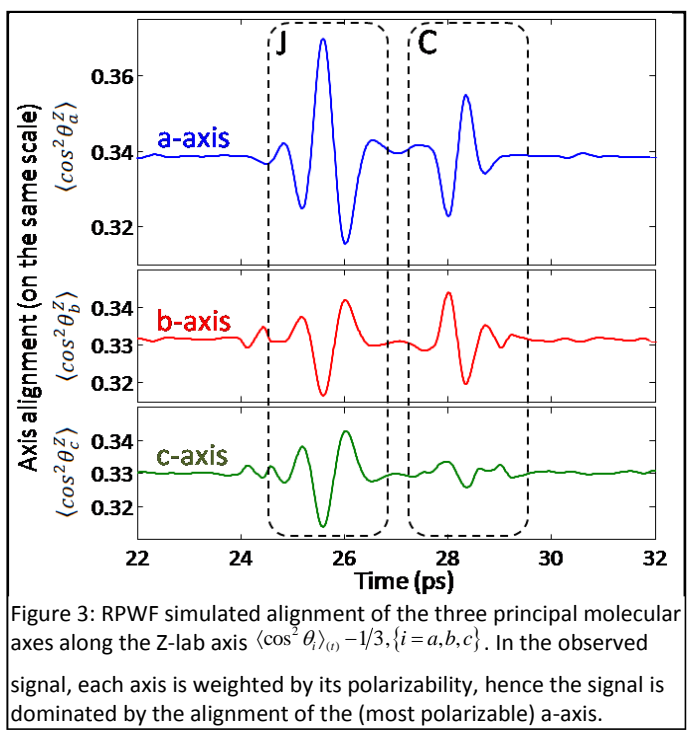

Figure 3: RPWF simulated alignment of the three principal molecular axes along the Z-lab axis $\langle\cos^2\theta_i\rangle_{(t)} - 1/3, \{i = a,b,c\}$. In the observed signal, each axis is weighted by its polarizability, hence the signal is dominated by the alignment of the (most polarizable) a-axis.

The J-type revivals involve the anisotropic distribution of all three molecular axes, with the molecular b- and c- axes anti-aligned at the instance of a-axis alignment and vice versa. This motion is driven via the most polarizable molecular axis (a-axis), forcing its alignment along the Z-lab axis, hence entails the anti-alignment of the b,c axes. Correspondingly the J-type revival period is given by 1/2[B+C], manifesting the rotation around the b and c molecular axes [Link to J-type revival cartoon]. Similarly, the C-type revivals describe rotation around the c-axis, with the a-axis aligned at the instance of b-axis anti-alignment and vice versa, hence the C-type revival period is given by 1/4C (not involving the B rotational coefficient) [Link to C-type revival cartoon] [44]. We note that another periodic feature, the A-type revivals with a period of 1/4A = 4.9ps is captured by the simulation for axes b and c, with opposite contributions of the two axes. However, due to very similar polarizabilities along the b- and c- axes, the weighted A-type signal is absent from the total alignment signal. (for detailed calculations of the alignment of the three molecular axes for the A-type revivals, see Fig.SI.4 in the SI section).

### THz-induced orientation of $SO_2$

We now turn to the THz-induced rotational dynamics of $SO_2$ at the pressure of 23torr and ambient temperature. The experimental setup used for THz-induced orientation measurements is a 4-f reflective setup similar to the one used in [29, for the experimental setup see SI.2]. An ultrashort laser pulse (100fs, 800nm) with 3.5mJ pulse energy was used to generate a ~3-uJ single-cycle THz pulse by tilted pulse front optical rectification [45]. The THz pulse is focused by an off-axis parabolic reflector into the center of a 10cm long static gas cell through 10mm-wide PTFE windows. The THz beam is re-collimated by an off-axis parabolic reflector at the output of the gas cell, and focused onto a 0.5mm long GaP crystal where it is electro-optically sampled by a variably delayed optical readout pulse [46].

The transmitted THz field consists of quazi-periodic THz-field bursts, emitted upon molecular orientation governed by the rotational dynamics of the excited sample. The emission of THz-fields however, can be understood by classical electrodynamics arguments: subsequent to their dipole-induced rotational motion, the $SO_2$ molecules exhibit transiently squeezed angular distributions. At these events, the permanent molecular dipoles are preferentially oriented toward the +z (or -z) lab-frame direction and give rise to the creation and annihilation of a macroscopic dipole in the sample. These oscillation of the macroscopic dipole are accompanied by electromagnetic field emission with a $\pi$-phase relative to the driving field, much like a resonantly driven antenna. The polarization of the antenna is $\pi/2$ phase shifted from that of the driving field, and the field emitted by the antenna is yet another $\pi/2$ shifted with respect to its polarization. In analogy, the transient orientation of the $SO_2$ dipoles (polarization of the medium) is $\pi/2$ phase shifted relative to the THz field carrier-envelope phase [29], and the emitted THz burst is another $\pi/2$ shifted with respect to the orientation.

Figure 4 shows the time-resolved signal emitted from a 23torr $SO_2$ sample contained in a 10cm long cell at room temperature (295K). The experimental data is depicted by the solid-blue line starting at ~5ps after the exciting THz field due to the large differences in amplitude of the obtained signal and the exciting THz field (with peak value of 0.87 in the scale of Fig. 4). The RPWF-simulated signal is depicted by the solid-red curve and is generated by simulating the orientation of the sample throughout the entire time domain and introducing a $\pi/2$ phase shift to provide the orientation-induced field-emission. The ensemble averaged orientation is calculated as the projection of the dipole vector (positive b-axis direction in the molecular frame) onto the lab *Z*-axis (the THz field polarization axis), $\langle \cos \theta_b^Z \rangle_{(t)}$. We note that the raw time-resolved electro-optic signal measured in the experiment consists of a large number of satellite signals emanating from various reflections of both the THz pump, and optical probe pulses in the GaP crystal, the vacuum cell PTFE windows and some residual water vapor residing in the THz beam path, outside the static gas cell. In order to retrieve the $SO_2$ signal we measured the response function of the optical system by EO sampling measurement with the gas cell evacuated and used it to deconvolve the $SO_2$ sample signal from the raw EO sampling data (see SI.3 for details).

Despite the crowded oscillatory character of the SO$_2$ emission in Fig.4, the fundamental revival features are clearly observed with a period of $1/(2A-B-C) \sim 9.8 ps$ corresponding to a hybrid-type revival, with $\Delta J = 0, \Delta K = 1$ previously termed by Felker [12]. The polarity of the signal alternates with the revival periods such that the phase of the even recurrence signals are $\pi$-shifted with respect to the odd ones (marked by 'DHyb' for dipole-induced hybrid type revivals ).

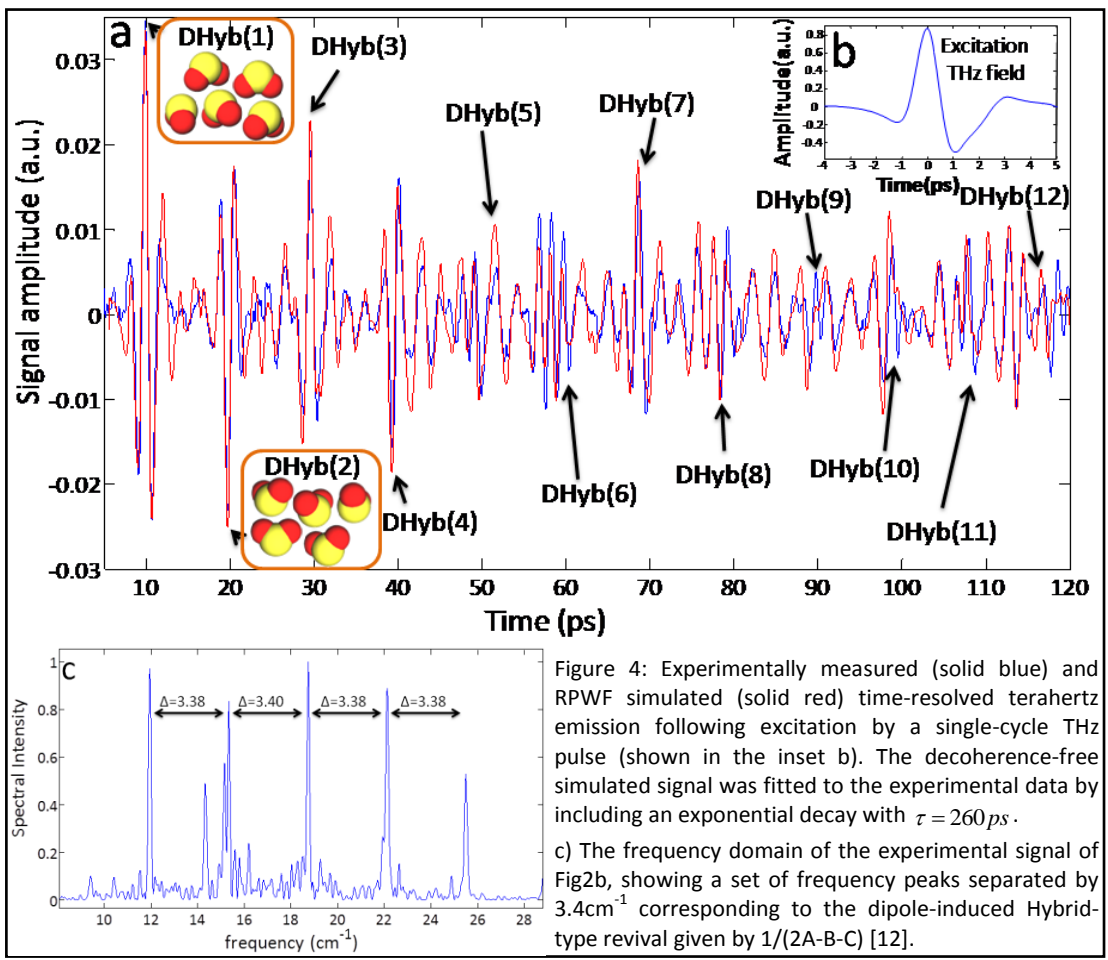

Figure 4: Experimentally measured (solid blue) and RPWF simulated (solid red) time-resolved terahertz emission following excitation by a single-cycle THz pulse (shown in the inset b). The decoherence-free simulated signal was fitted to the experimental data by including an exponential decay with $\tau = 260 ps$.

c) The frequency domain of the experimental signal of Fig2b, showing a set of frequency peaks separated by 3.4cm$^{-1}$ corresponding to the dipole-induced Hybrid-type revival given by 1/(2A-B-C) [12].

The existence of the DHyb revivals as the main feature of Fig.4a is apparent from its Fourier transform shown in Fig.4c manifesting several peaks distant by $3.4 cm^{-1} = 2A - B - C$.

Most important to achieving control over molecular angular distributions is the association of the recurring signals with the orientation of the molecular dipoles - during these events, the molecular dipoles preferentially point toward the positive or negative z-axis direction (see a sketch of SO$_2$ dipole directions for the first odd and even DHyb revivals in Fig.4). The Dhyb revivals persist for long times under field-free conditions, enabling additional interactions with laser pulses for enhanced rotational control or non-linear spectroscopic interrogation of the molecules in their molecular frame in addition to the periodically emitted THz signals.

Note that the periodic orientation revivals of Fig.4 (with 9.8ps period) are absent from the alignment results of Fig.2 and vice versa, the J-type (26.1ps) and C-type (28.4ps) revivals are absent from the THz-induced dynamics of Fig.4. Namely, the THz and optical pulses induce distinct rotational dynamics in asymmetric $SO_2$ molecules, and induce selective rotational motions due to the different nature of their interactions, mediated via the permanent dipole and the anisotropic polarizability respectively. Unlike linear and symmetric top molecules, where the dipole and the most polarizable axis are parallel, in $SO_2$, as in most asymmetric molecules, the dipole and most polarizable axes are not parallel to one another (In $SO_2$ they are situated perpendicularly). This manifests in distinctively different dynamics induced the THz and the optical fields, providing two distinct handles over the rotational dynamics of asymmetric molecules. Well concerted utilization of these handles pave a path to realizing three-dimensional control over molecular angular distributions, inducing torsion in molecules and can aid in the breaking-down congested rotational dynamics by associating the rotational responses induced by THz and optical fields separately to their molecular-frame handles. Co-application of THz and optical rotational handles can significantly extend the currently available rotational control playground to yield novel rotational responses and dynamics unreachable by either of these fields alone.

We wish to thank Ronnie Kosloff (HUJI) for stimulating discussions. This work is supported by the Binational Science Foundation grant no. 2012021 (S.K) and by CMST COST Action CM1405 MOLIM (S. K), by the Israeli Science Foundation grant no.1065/14  (S.F.), the Marie Curie CIG grant no. 631628 (S.F) and in part by the Wolfson family grant (S.F.)

† Email: shimshonkallush@braude.ac.il
‡ Email: sharlyf@post.tau.ac.il